\documentclass[twocolumn,aps,superscriptaddress,prl]{revtex4}
\usepackage{graphicx}
\usepackage{times}
\usepackage{amsmath}
\usepackage{amssymb}
\usepackage{units}
\usepackage{stmaryrd} 
\DeclareGraphicsExtensions{.pdf,.png, .eps ,.jpg}

\begin{document}
\preprint{0}

\title{Evidence of reduced surface electron-phonon scattering in the conduction band of $\mathbf{Bi_{2}Se_{3}}$ by non-equilibrium ARPES}

\author{A. Crepaldi}
\email{alberto.crepaldi@elettra.trieste.it}
\affiliation{Institute of Condensed Matter Physics (ICMP), \'Ecole Polytechnique
F\'ed\'erale de Lausanne (EPFL), CH-1015 Lausanne,
Switzerland}
\affiliation{Elettra - Sincrotrone Trieste, Strada Statale 14 km 163.5 Trieste, Italy}

\author{F. Cilento}
\affiliation{Elettra - Sincrotrone Trieste, Strada Statale 14 km 163.5 Trieste, Italy}

\author{B. Ressel}
\affiliation{Elettra - Sincrotrone Trieste, Strada Statale 14 km 163.5 Trieste, Italy}
\affiliation{University of Nova Gorica, Vipavska 11 c, 5270 Ajdov\v{s}\v{c}ina, Slovenia}

\author{C. Cacho}
\affiliation{Central Laser Facility, STFC Rutherford Appleton Laboratory, Harwell, United Kingdom}

\author{J. C. Johannsen}
\affiliation{Institute of Condensed Matter Physics (ICMP), \'Ecole Polytechnique
F\'ed\'erale de Lausanne (EPFL), CH-1015 Lausanne, Switzerland}

\author{M. Zacchigna}
\affiliation{C.N.R. - I.O.M., Strada Statale 14 km 163.5 Trieste, Italy}

\author{H. Berger}
\affiliation{Institute of Condensed Matter Physics (ICMP), \'Ecole Polytechnique
F\'ed\'erale de Lausanne (EPFL), CH-1015 Lausanne,
Switzerland}

\author{Ph. Bugnon}
\affiliation{Institute of Condensed Matter Physics (ICMP), \'Ecole Polytechnique
F\'ed\'erale de Lausanne (EPFL), CH-1015 Lausanne,
Switzerland}

\author{C. Grazioli}
\affiliation{University of Nova Gorica, Vipavska 11 c, 5270 Ajdov\v{s}\v{c}ina, Slovenia}

\author{I. C. E. Turcu}
\affiliation{Central Laser Facility, STFC Rutherford Appleton Laboratory, Harwell, United Kingdom}

\author{E. Springate}
\affiliation{Central Laser Facility, STFC Rutherford Appleton Laboratory, Harwell, United Kingdom}

\author{K. Kern}\affiliation{Institute of Condensed Matter Physics (ICMP), \'Ecole Polytechnique
F\'ed\'erale de Lausanne (EPFL), CH-1015 Lausanne,
Switzerland}
\affiliation{Max-Plank-Institut f\"ur Festk\"orperforschung,
D-70569, Stuttgart, Germany}

\author{M. Grioni}
\affiliation{Institute of Condensed Matter Physics (ICMP), \'Ecole Polytechnique
F\'ed\'erale de Lausanne (EPFL), CH-1015 Lausanne,
Switzerland}

\author{F. Parmigiani}
\affiliation{Elettra - Sincrotrone Trieste, Strada Statale 14 km 163.5 Trieste, Italy}
\affiliation{Universit\`a degli Studi di Trieste - Via A. Valerio 2 Trieste, Italy}

\date{\today}

\begin{abstract}

The nature of the Dirac quasiparticles in topological insulators calls for a direct investigation of the electron-phonon scattering at the \emph{surface}. By comparing time-resolved ARPES measurements of the TI $\mathrm{Bi_{2}Se_{3}}$ with different probing depths we show that the relaxation dynamics of the electronic temperature of the conduction band is much slower at the surface than in the bulk. This observation suggests that surface phonons are less effective in cooling the electron gas in the conduction band.

\end{abstract}

\maketitle


The scientific and technological interest on topological insulators (TIs) stems from the unusual properties of their topologically protected metallic surface states, which exhibit a linear dispersion and a characteristic spin helicity \cite{FU_PRL_2007, Hsieh_Nature_2008, Hasan_RMP_2010,Xia_NatPhys_2009, Chen_Science_2009, Zhang_NatPhys_2009, Zhao_NLett_2011}. For the Dirac quasiparticles elastic backscattering is forbidden by time-reversal symmetry, and transport is controlled by scattering events mediated by phonons. 
Attempts to measure the strength of the electron-phonon coupling in the representative TI $\mathrm{Bi_{2}Se_{3}}$ by angle-resolved photoelectron spectroscopy (ARPES) have produced somewhat conflicting results. The estimated values of the dimensionless coupling constant 
$\lambda$ vary from small ($\lambda~\sim~0.08$) \cite{Valla_Pan_PRL_2012} to moderate ($\lambda~\sim~0.25 $) \cite{Bianchi_PRB_2011}. Time-resolved ARPES (tr-ARPES) can tackle the problem in the time domain, complementary to the energy domain of conventional ARPES at equilibrium \cite{Sobota_PRL_2012, Perfetti_arxiv_2012, Gedik_PRL_2012, Crepaldi_2012}. In pump-probe tr-ARPES experiments, the electrons excited by a light pulse are described by an effective Fermi-Dirac (FD) distribution. The relaxation of the electronic temperature ($T_{e}$), as well as the variation of the chemical potential ($\mu$) that reflects photo-doping of the conduction band, provide fundamental information on the de-excitation mechanisms, namely between the conduction band (CB) and the topologically protected surface state \cite{Gedik_PRL_2012, Crepaldi_2012}. In a recent experiment on $\mathrm{Bi_{2}Se_{3}}$, the contribution of various phonon modes to the electronic cooling has been addressed by comparing the relaxation dynamics of the FD distribution at various sample temperatures and for different charge densities \cite{Gedik_PRL_2012}. 


\begin{figure*}[ttt]
 
  \includegraphics[width = 0.9 \textwidth]{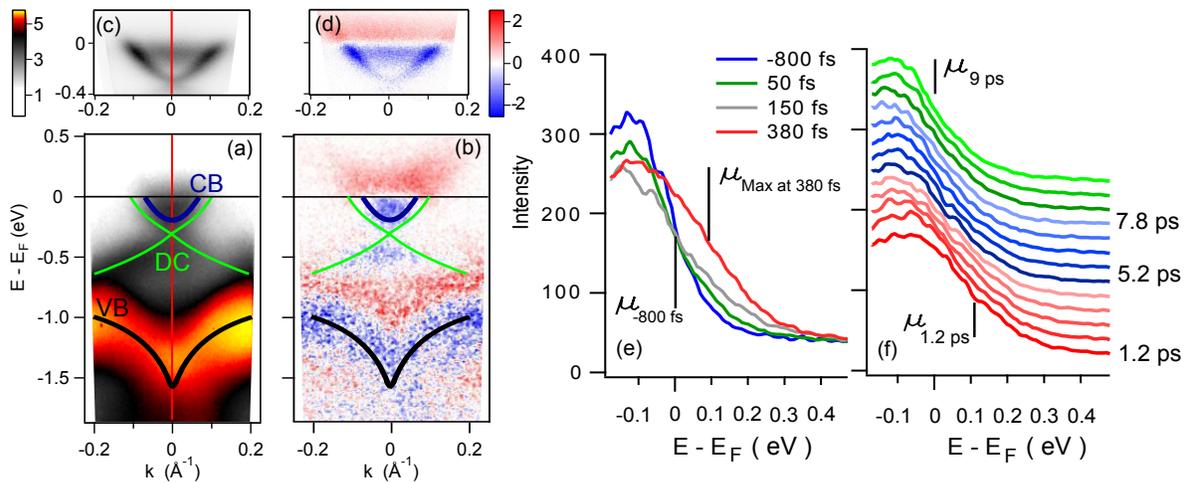}
  \caption[didascalia]{(Color online) (a) ARPES image of the band structure of $\mathrm{Bi_{2}Se_{3}}$ around the $\Gamma$ point, measured with the surface sensitive probe (17.5 eV) before ($-500$~fs) the optical excitation. Blue (green) lines outline the dispersion of the bulk (surface) states. (b) tr-ARPES image obtained as difference between the ARPES data at positive delay times ($+ 300$~fs) and the reference signal before optical excitation ($- 500$~fs) with \emph{s}-polarized light at 1.59 eV. (c, d) band structure and tr-ARPES image as panel (a, b) measured with 6.2 eV photon energy. (e, f) evolution of the energy distribution curves (EDCs) taken from the ARPES data at the $\Gamma$ point at selected delay times in the two temporal regions t $<$ 0.5 ps (c) and t $>$ 1 ps (d).   
   }
  \label{fig:arpes1}
\end{figure*}


In this work we present a tr-ARPES investigation of the conduction band dynamics in $\mathrm{Bi_{2}Se_{3}}$, where two different photon energies are exploited to vary the surface sensitivity.  
Standard tr-ARPES experiments, performed with laser-based sources at 6.2 eV photon energy, are rather bulk sensitive, due to the very low kinetic energy of the photo-electrons \cite{Okawa_PRB_2009}. The comparison between \emph{more} bulk sensitive ($h\nu$~=~6.2 eV; UV) and \emph{more} surface sensitive ($h\nu$~=~17.5 eV; extreme UV, EUV) measurements reveals two different relaxation dynamics for $T_{e}$ in the conduction band. Namely, we observe a \emph{freezing} of $T_{e}$ to an elevated value ($\sim 600$ K) at the surface but not in the bulk, suggesting a reduced efficiency of the phonons in the electronic cooling \emph{at the surface}. 

A quantitative estimation of the photo-electron mean free path, $l$, as a function of the photo-electron kinetic energy is challenging. In particular, it is well established that at low kinetic energies ($<$ 5 eV), the photo-electron escape depth is strongly material dependent \cite{Lindau_1974, Mao_APL_2008}. An estimation of $l$ for $\mathrm{Bi_{2}Se_{3}}$ gives $\sim$~2 - 3 nm at 6.2 eV photon energy \cite{Perfetti_arxiv_2012}. This corresponds to $\sim$~3 - 4 quintuple layers (QL). Whereas $l$ $<$ 1 nm at 17.5 eV  photon energy, hence the escape depth is less than 1 QL  \cite{Lindau_1974, Mao_APL_2008, Hufner_book}.  
For $\mathrm{Bi_{2}Te_{3}}$ \cite{Chulkov_2012},  $\mathrm{Bi_{2}Se_{3}}$ \cite{Ye_arxiv_2011, Zhu_PRL_2012, Bahramy_Ncom_2012} and other TI system like $\mathrm{PbBi_{2}Te_{4}}$ and  $\mathrm{PbBi_{2}Te_{7}}$ \cite{Eremeev_Ncom_2012}, the actual models suggest that the surface state wave function receives a large (10-20 $\%$) contribution form the second QL. This implies that the Dirac particles extend also below the surface  (second and third QLs),  and in principle they can be fully probed with a suitable photon energy. 


The EUV tr-ARPES experiments were performed at the Artemis facility at the Central Laser Facility of the Rutherford Appleton Laboratory. The laser output at 780 nm and 1 kHz repetition rate was split in two beams. One was used to generate the tunable IR pump (here we show the results for optical excitation at 1.59 eV, and similar results were obtained by pumping at 0.95 eV \cite{suppl_1}); the other to generate the tunable high-harmonic (HH) EUV probe. We selected the 11th harmonic at $\sim$~17.5 eV. The overall time resolution was 60 fs and the energy resolution was 180 meV  \cite{suppl_1, Petersen_PRL_2011, Poletto_OE_2011, Bauer_Nature_2011}. The tr-ARPES experiments with 6.2 eV probe energy were performed at the T-ReX laboratory, Elettra (Trieste), with a Ti:Sa regenerative amplifier producing 780 nm laser pulses at 250 kHz repetition rate. Electrons were photoemitted by the fourth harmonics of the laser source (6.2 eV) obtained by harmonic generation in phase-matched BBO crystals. In this setup the energy resolution was 10 meV and the temporal resolution 300 fs. Particular attention was paid in order to match the experimental conditions in the two different setups. In particular, we used the same pump photon energy (1.59 eV) with fluence equal to 160~$\pm$~30~$\mu J/cm^{2}$. The high quality n-doped $\mathrm{Bi_{2}Se_{3}}$ samples were taken from the same batch. They were cleaved under UHV ($<$ 5 10$^{-10}$~mbar) at room temperature. In each experiment the probe beam intensity was chosen in order to make negligible space charge effects, as witnessed by checking the energy and momentum resolution of the measured ARPES spectra.

\begin{figure}[th]
  \includegraphics[width = 0.45 \textwidth]{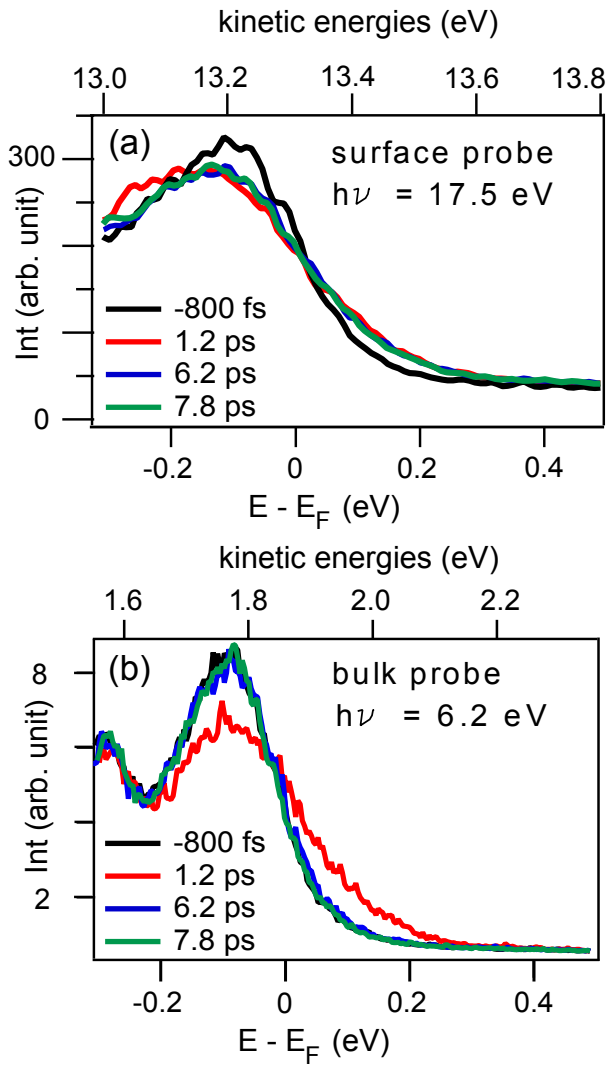}
  \caption[didascalia]{(Color online) EDCs at the $\Gamma$ point for different delay times: (a) surface sensitive probe (17.5 eV); (b) bulk sensitive probe (6.2 eV). The different broadening at negative delay time between (a) and (b) reflects the different energy resolution of the two set-ups. The broadening at positive delay times reflects instead an increase of $T_{e}$. In (a) the EDCs are shifted in energy to compensate the variations of $\mu$ (Fig. 1).  The three EDCS at positive delays overlap, suggesting that $T_{e}$ reaches a steady value. In (b), by contrast, at long time delays ($t>$ 5 ps) $T_{e}$ has already relaxed back to room temperature.  
  }
  \label{fig:arpes2}
\end{figure}


Figure~1 compares the results obtained with $17.5$ eV and $6.2$ eV photon energies. Figure~1~(a) shows as a reference the electronic states of $\mathrm{Bi_{2}Se_{3}}$ around the $\Gamma$ point before the optical excitation. In this and in the following figures, the binding energy scale always refers to the Fermi level position at negative delay time, i.e. before the optical perturbation. The Dirac cone (DC) disperses across the band-gap connecting the V-shaped valence band (VB) to the conduction band (CB) and it intersects the Fermi level ($E_{F}$) at $k_{F} = \pm 0.1~\mathring{A}^{-1}$. The data are consistent with previous ARPES experiments \cite{Xia_NatPhys_2009, Hsieh_nature_2009,Bianchi_ACS_2012}, but the full-bandwidth preservation of the HH pulse prevents us from discerning the DC from the CB. On the other hand, the higher surface sensitivity of the EUV light enables us to investigate, for the first time, the out-of-equilibrium dynamics of the CB \emph{at the surface}.

Figure~1~(b) shows a tr-ARPES image obtained as a difference between the ARPES data at positive delay times ($+ 300$~fs) and the reference signal before optical excitation ($- 500$~fs). The signal reflects a modification of the FD distribution \cite{Sobota_PRL_2012, Perfetti_arxiv_2012, Gedik_PRL_2012, Crepaldi_2012}. Below $E_{F}$, the DC and CB drastically lose spectral weight, which is transferred above $E_{F}$. Similarly, we interpret the weak variation of intensity in the VB as an effect of thermal broadening of the valence band due to the pump pulse \cite{Othonos_JApp_1998}. Figure~1~(c - d) show the same as panel (a - b) but as measured with 6.2 eV photon energy. In the two datasets the relative intensities of the surface state and the CB are different owing to matrix element effects, which provide information on the transition probability. Once the spectral features are energetically resolved, their relaxation dynamics in tr-ARPES are not affected by matrix element effects. In the present experiments the CB is clearly resolved for both the probe photon energies, allowing us to investigate its non-equilibrium dynamics.

We now focus on the evolution of the FD distribution describing the \emph{hot} electrons in the CB after the optical perturbation.  Figure~1~(e - f) show the evolution of the energy distribution curves (EDCs) at the $\Gamma$ point (as indicated by the red line in panel (a) and (c)), integrated over a small ($\pm0.025~\mathring{A}^{-1}$) wave vector window, as a function of the delay time between the $1.59$ eV pump and the $17.5$ eV probe pulses. The small integration window ensures that only the dynamics of the CB is probed. The evolution of the EDCs clearly indicates the existence of two different dynamics, for $t < 0.5$~ps (Fig.~1~(c)) and $t > 1$~ps (Fig.~1~(d)). At short delay times a fast broadening is observed, which is attributed to the rapid increase of $T_{e}$. The thermal broadening of the FD function reaches its maximum at $t \sim 150$~fs (gray EDC in Fig.~1~(c)). During this fast process the position of the chemical potential $\mu$ is unaltered. Only during the first part of the relaxation of $T_{e}$ we do observe a large shift of $\mu$ (red EDC in Fig.~1~(c)). Figure~1~(d) shows the evolution of the EDCs between 1.2 ps and 9 ps. Interestingly, while $\mu$ recovers its equilibrium value during this time interval, $T_{e}$ does not, as shown by the persistent broadening of the EDCs. 


A second set of data measured with $6.2$ eV photons enables a direct comparison of the relaxation dynamics at the surface and in the bulk of $\mathrm{Bi_{2}Se_{3}}$. Selected surface- and bulk-sensitive spectra are shown in Fig. 2~(a) and (b), respectively. Each panel displays four EDCs measured at the $\Gamma$ point at -0.8 ps, 1.2 ps, 6.2 ps and 7.8 ps delay times. The binding energy window is the same in (a) and (b), but the kinetic energy of the photoelectrons varies by a factor $\sim~10$ (see the top energy scale). The different widths of the EDCs in (a) and (b) at negative delay time is due to the different energy resolution of the two experimental set-ups. The broadening at positive delay times reflects the increase in $T_{e}$. In (a) the EDCs are rigidly shifted along the energy axis in order to compensate changes in $\mu$. The three EDCs at positive delays overlap, thus indicating the formation of a state with large $T_{e}$, which lasts several picoseconds. By contrast, in (b) at the larger delay times ($>$ 5 ps) the bulk-sensitive EDCs coincide with the one at negative delay. Clearly, a nearly steady state with a high electronic temperature is generated at the surface, but not in the bulk. 


We performed a quantitative analysis of the evolution of $T_{e}$ and $\mu$ by fitting the leading edge of the EDCs -- the so called electronic \emph{hot tail} -- with a FD function, convolved with a Gaussian to account for the finite energy resolution. This simple fitting procedure neglects the details contained in the analytical form of the density of states, which have been recently analyzed in other high resolution laser tr-ARPES experiments \cite{Crepaldi_2012, Gedik_PRL_2012}. Nevertheless, it provides us with a quantitative description of the peculiar effect, already visible in the raw data of Fig. 2~(a). Figure~3~(a - b) display the best fit parameters for $\Delta\mu$, defined as the variation of the chemical potential before and after optical excitation, and $T_{e}$, as a function of the delay time. Blue and red markers indicate respectively the results for the surface and bulk sensitive probe. In the bulk, the relaxation of $T_{e}$ is fitted with a single exponential decay with $\tau_{T} = 2.7$~ps, times a step function to reproduce the rise time convolved with a Gaussian to account for the temporal resolution (green line in panel (b)), in good agreement with the literature \cite{Crepaldi_2012, Sobota_PRL_2012}.  At the surface, a fit of $T_{e}$ (black line in panel (b)) is achieved with a single exponential decay plus a constant to account for the dynamics at large delay times. This function is multiplied by a step function to reproduce the rise time, convolved with a Gaussian to account for the temporal resolution. The high temporal resolution of the EUV setup enables us to resolve a fast relaxation with a characteristic decay time $\tau_{T} = 160$~fs. The underlying de-excitation mechanism is responsible also for the increase in $\Delta\mu$, resulting from the variation in the electron density in the lower energy branch of the CB. This indicates that electrons populate the bottom of the CB after scattering from higher energy states, as proposed in Ref. \onlinecite{Sobota_PRL_2012}.

The second dynamics observed in $T_{e}$ is approximated by a constant at large delay times, and the resulting best fit parameter is $T_{0}= 640$ K. A black arrow indicates the crossover between the two relaxation dynamics. The same delay time in Fig.~3~(a) corresponds to the maximum of $\Delta\mu$. The two relaxation dynamics are unaffected by changing the pump energy to $0.95$ eV. Also at this excitation energy $T_{e}$ does not relax back to its equilibrium value, but it exhibits a \emph{plateau} at $\sim$ 550 [suppl. material]. 


Previous bulk sensitive tr-ARPES studies reported a characteristic relaxation time of $T_{e}$ of the order of few ps (2.5 ps \cite{Crepaldi_2012}, 1.7 ps \cite{Sobota_PRL_2012}), in agreement with our bulk sensitive probe. In all the previous investigations such timescale was unambiguously attributed to the electron-phonon interaction. The energy deposited in the electronic bath by the pump pulse is transferred to the lattice via scattering with phonons. The characteristic relaxation time of $T_{e}$ is a measure of the strength of the electron-phonon scattering in the material \cite{Allen_PRB_1987}. We interpret the long lasting out-of-equilibrium value of $T_{e}$ in the conduction band of $\mathrm{Bi_{2}Se_{3}}$ as a manifestation of a reduced efficiency of the phonon scattering \emph{at the surface}, as a mechanism to remove energy from the excited electronic bath. 
Interestingly, the temperature characterizing the \emph{plateau} in Fig.~3~(a) ($T_{0} \sim 640$~K) is comparable to the temperature ($600$~K) where the cooling by optical phonons is expected to be less effective for a Dirac particle \cite{Gedik_PRL_2012}, while a different electron density dependent relaxation mechanism dominates \cite{Gedik_PRL_2012}. Our data suggest that below $640$~K the latter scattering process is strongly suppressed at the surface influencing the relaxation dynamics of the CB. The nature of this second scattering mechanism is still under debate and further studies are required to address the role of surface acoustic phonons. It would be of importance also to clarify whether this effect is related simply to the surface properties of the material, owing to the broken translational symmetry and the lower coordination number, or whether it reflects some more fundamental properties related to the topological protection of the Dirac particle in $\mathrm{Bi_{2}Se_{3}}$. The relaxation dynamics in CB is influenced by the Dirac particles dynamics through the scattering mechanisms involving transitions among the different states, modeled by a system of rate equations as proposed by Hajlaoui et al.\cite{Perfetti_arxiv_2012}.

The relaxation of $\Delta\mu$ can result both from the electron-phonon scattering and from the diffusion of the photoexcited electrons, which leaves the probed region, thus contributing to recover the charge density in CB at equilibrium. The difference between the relaxation of $\Delta\mu$ at the surface and in the bulk suggests that the involved mechanisms must vary significantly as a function of the photo-electron escape depth. We observe that in the bulk, $\Delta\mu$ and $T_{e}$ decay on similar timescales. The electron-phonon scattering has been proposed to be the most important contribution in the relaxation of the $T_{e}$ \cite{Gedik_PRL_2012, Crepaldi_2012}. Hence, these similar relaxation times suggest that the electron-phonon scattering is also the dominant mechanism governing the dynamics of $\Delta\mu$. Conversely, at the surface the electron-phonon scattering is reduced and the relaxation of $\Delta\mu$ is governed by the diffusion effects. Therefore, the relaxation becomes slower at the surface, and it cannot be modeled by an exponential decay.


In our surface-sensitive experiment we also investigated the dependence of the non-equilibrium electronic dynamics on the absorbed pump fluence, which we varied between 50 $\mu J/ cm^{2}$ and 400 $\mu J/ cm^{2}$ (at 1.59 eV). At the ps timescale, we observe (Fig.~3~(c)) the formation of a plateau in the relaxation of $T_{e}$, which is essentially independent of the pump fluence ($T_{0}$ $\sim$ 600 K). This confirms that the persistent non-equilibrium $T_{e}$ is not an artifact due to the average heating of the crystal, which is expected to scale with the adsorbed fluence. Furthermore this observation indicates that the plateau is not an artifact due to pump-induced space charge effects, as the measured energy broadening of the Fermi Dirac distribution does not depend on the pump fluence. The  almost linear dependence of $\Delta\mu$ on the pump fluence (Figure~3~(d)) reflects the larger density of excited charges in CB.


\begin{figure}[t]
  \includegraphics[width = 0.45\textwidth]{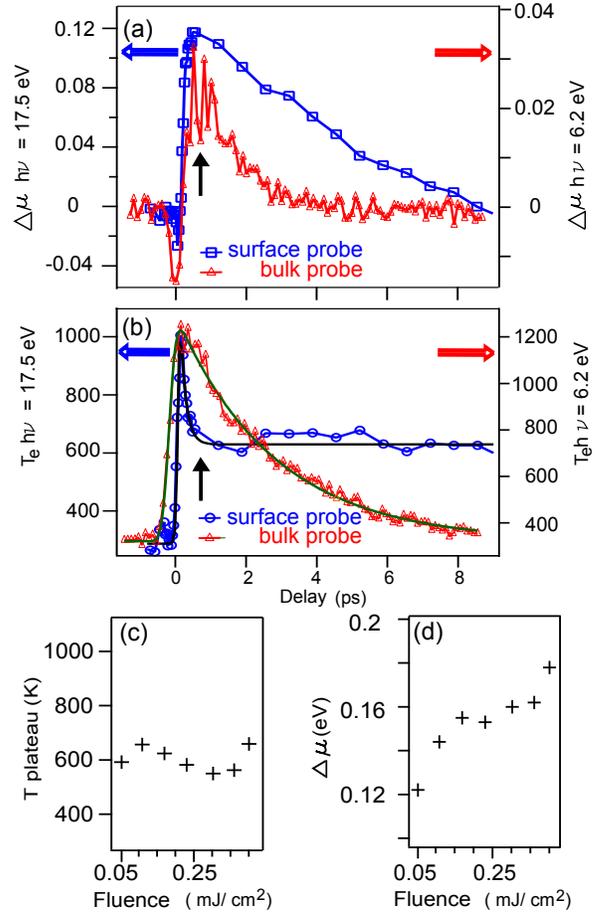}
  \caption[didascalia]{ (Color online) (a, b) $\mu$ and $T_{e}$ as determined by fitting a Fermi-Dirac function to the EDCs at the $\Gamma$ point shown in Fig.2. Blue (red) markers refer to 17.5 eV (6.2 eV) probe energy. In (b) the black solid line shows the best fit obtained with a single decay exponential plus a constant, with characteristic relaxation time $\tau_{T} = 160$~fs. At large delay times a steady state with $T_0=640$ K is achieved. The pump fluence dependence of the maximum change of $\mu$ (d) and of the plateau value of $T_{e}$ (c), probed with $17.5$ eV photons, is reported.
 }
  \label{fig:arpes3}
\end{figure}


In summary, we have performed a probe energy dependent tr-ARPES study of the relaxation dynamics of the electronic temperature of the conduction band in $\mathrm{Bi_{2}Se_{3}}$.
The use of 6.2 eV and 17.5 eV probe energies, thanks to their different surface sensitivity, offers a unique possibility to investigate the electron-phonon scattering mechanisms \emph{at the surface} of a TI. 
We observe two distinct dynamics in the relaxation of $T_{e}$ at the surface. An ultrafast decrease with $\tau_{T} \sim 160$ fs is followed by the formation of a steady state with a value $T_{0} \sim 600$~K which lasts several ps. By contrast, in the bulk, $T_{e}$ relaxes back to its equilibrium value with $\tau_{T} \sim 2.7$ ps. The nearly steady value of $T_{e}$ indicates that the electron-phonon scattering is reduced at the surface of $\mathrm{Bi_{2}Se_{3}}$. It would be important to verify whether a similar reduction of the electron-phonon scattering affects also the topologically protected surface state. In fact, the surface state wave function extends in the sub-surface region, \emph{i.e.} in the second QL and more weakly in the third QL. At 6.2 eV photon energy these regions are within the primary photo-electrons probed volume. Whereas, they are only partially  accessible  at 17.5 eV photon energy. The possibility of a reduced electron-phonon scattering in the surface state of $\mathrm{Bi_{2}Se_{3}}$ is supported by the experimental observation of a weak electron-phonon coupling ($\lambda = 0.08$) of the surface state, as recently reported by conventional surface sensitive ARPES \cite{Valla_Pan_PRL_2012}.  The reduced electronic cooling at the surface of TIs would have strong implications for applications of $\mathrm{Bi_{2}Se_{3}}$ in (opto-) spintronics devices.
We believe that a detailed description of the phonon scattering processes at the picoseconds timescale must adequately take into account the observed different behavior of the surface and the bulk of $\mathrm{Bi_{2}Se_{3}}$.

We acknowledge technical support from Phil Rice. Access to STCF's Artemis facility was funded by LASERLAB-EUROPE II (grant n° 228334). This work was supported in part by the Italian Ministry of University and Research under Grant Nos. FIRBRBAP045JF2 and FIRB-RBAP06AWK3 and by the European Community–Research Infrastructure Action under the FP6 "Structuring the European Research Area" Programme through the Integrated Infrastructure Initiative "Integrating
Activity on Synchrotron and Free Electron Laser Science" Contract No. RII3-CT-2004-506008.


\end{document}